\documentclass[%
 aps,pra,reprint,
 superscriptaddress, 
 showpacs,showkeys,
 noeprint,nopreprintnumbers,
 amsmath,amssymb,
 longbibliography,
 nofootinbib,
 floatfix
]{revtex4-1}
\usepackage[utf8]{inputenc}
\usepackage[pdftex]{graphicx}
\usepackage[usenames, dvipsnames]{xcolor}

\usepackage{braket} 
\usepackage[normalem]{ulem} 
\usepackage{dcolumn}
\usepackage{bm}
\usepackage{bbold} 
\usepackage{hyperref} 
\usepackage{comment}
\usepackage{bibunits}
\usepackage{ dsfont }
\usepackage{ mathrsfs }
\usepackage{cancel}
\usepackage{xparse}
\usepackage{physics}
\usepackage{soul}
\usepackage{ gensymb }
\usepackage{bibunits}
\defaultbibliographystyle{apsrev4-1}
\defaultbibliography{main-arxiv}

\graphicspath{ {./IMG/} {./} }
\newcommand{\ksig}{\ensuremath{{\bf k}\sigma}}

\newcommand{\fir}[1]{Fig.~\ref{#1}}

\newcommand{\eqr}[1]{Eq.~(\ref{#1})}

\newcommand{\ee}{\textrm{e}}

\begin{document}

\def\mytitle{Higgs mode stabilization by photo-induced long-range interactions in a superconductor
}
\title{\mytitle}

\author{Hongmin Gao}
 \email{hongmin.gao@physics.ox.ac.uk}
 \affiliation{Clarendon Laboratory, University of Oxford, Parks Road, Oxford OX1 3PU, United Kingdom}

\author{Frank Schlawin}
 \email{frank.schlawin@mpsd.mpg.de}
 \affiliation{Max Planck Institute for the Structure and Dynamics of Matter, Hamburg, Germany}

\author{Dieter Jaksch}
 \email{dieter.jaksch@physics.ox.ac.uk}
 \affiliation{Clarendon Laboratory, University of Oxford, Parks Road, Oxford OX1 3PU, United Kingdom}

\date{\today}

\begin{abstract}
We show that low-lying excitations of a 2D BCS superconductor are significantly altered when coupled to an externally driven cavity, which induces controllable long-range attractive interactions between the electrons. 
We find that they combine non-linearly with intrinsic local interactions to increase the Bogoliubov quasiparticle excitation energies, thus enlarging the superconducting gap. 
The long-range nature of the driven-cavity-induced attraction qualitatively changes the collective excitations of the superconductor. Specifically, they lead to the appearance of additional collective excitations of the excitonic modes. 
Furthermore, the Higgs mode is pushed into the gap and now lies below the Bogoliubov quasiparticle continuum such that it cannot decay into quasiparticles. This way, the Higgs mode's lifetime is greatly enhanced. 
\end{abstract}

\maketitle

\begin{bibunit}

\nocite{apsrev41Control}

\section*{Introduction}
Low-lying excitations in quantum systems are important as they play a major role in determining the macroscopic orders and the microscopic dynamics. 
Superconductors, in particular, host a range of quasiparticle and collective excitations. They dictate properties such as superconducting critical temperatures, electromagnetic responses \cite{Parks_sc,Sun2020} and topological orders \cite{QiRMP2011,Sato_2017}; they carry information on the ground state gap properties \cite{Schwarz2020,Krull2016}; and they also govern the out-of-equilibrium dynamics of the superconductors \cite{Chang1978,BasovRMP2005,Dienst2011,Matsunaga13, Matsunaga14,Laplace2016,dolgirev2021periodic,Vaswani2021}. 

In a conventional, Bardeen–Cooper–Schrieffer (BCS) superconductor, 
the electron attraction is local. Electrons form Cooper pairs and condense into a superfluid described by a complex order parameter below a critical temperature \cite{TinkhamSuperconductivity}. 
Cooper-pair breaking leads to Bogoliubov quasiparticle excitations which are gapped near the Fermi surface. 
Additionally, 
the spontaneous breaking of the U(1) symmetry upon condensation creates a Goldstone (phase) mode \cite{Goldstone1961}. Due to the coupling to the electromagnetic fields, this collective mode is absorbed by the fields through the Anderson-Higgs mechanism \cite{Anderson63,Higgs1964}. Its manifestation as plasmon oscillations in a 2D superconductor was observed in, e.g., Ref.~\cite{Dunmore95}. 
Alongside the Goldstone mode, an orthogonal Higgs (amplitude) mode \cite{Shimano20} with energy twice the gap size, i.e., on the edge of the quasiparticle excitation continuum, is also created. This collective mode is an analogy (arguably the only one, based on the criterion of local gauge-invariance \cite{Pekker15}) in condensed matter systems to the Higgs boson in particle physics discovered in 2012 \cite{ATLASCollaboration}.

Superconductors with subdominant pairing instabilities exhibit another type of collective excitations termed excitonic/Bardasis-Schrieffer (BS) modes \cite{Bardasis1961}. 
They are fluctuations of a superconducting order parameter in the subdominant channels 
\cite{Allocca19,Sun2020}. 
Excitonic modes are stable collective excitations with energies below the quasiparticle excitation continuum and decreasing to zero as the subdominant pairing strengths approach the dominant one. 

The excitonic and, to an even larger extent, the Higgs modes have received considerable interest since their theoretical predictions. 
This interest not only stems from their direct link to spontaneous symmetry breaking, but is also due to the fact that measurements on them can reveal ground state gap symmetries and multiplicities \cite{Krull2016,Schwarz2020}, couplings to other collective modes \cite{Chu2020}, and information on the electronic interactions present \cite{Murakami16,Kretzschmar2013,Bohm2014,Jost2018}. 
However, these collective modes have proven to be rather elusive to experimental detection. 
This is partly due to the fact that they do not couple strongly to the electromagnetic fields \cite{Sun2020}: particle-hole symmetry prevents linear coupling to the Higgs mode \cite{Shimano20,TsujiPRB15}, while the lack of a dipole moment and optical selection rules forbid far-field optical excitations of the excitonic modes \cite{Sun2020}. 
Theoretical proposals have suggested inducing linear coupling by applying supercurrents \cite{Allocca19,Raines20}, or detecting these modes with cryogenic near-field nano-optics \cite{BasovRMP14,Lundeberg17,Dias18,Ni2018,Wang2020}. 

Aside from the issue of weak coupling to light, 
the excitonic modes, though stable, also require a strong subdominant order to be well-separated from the continuum. As a result, they have been detected only recently by Raman spectroscopy in iron-based superconductors \cite{Kretzschmar2013,Bohm2014,Jost2018}. 

For the Higgs mode, there is the difficulty that it usually decays rapidly 
into quasiparticle-quasihole pairs, even in the low temperature limit where the electron relaxation is much slower than the dynamics of the superconductor \cite{Volkov73,Pekker15}.
So far, exceptions to this have been found only in superconductors with strong disorder \cite{Sherman2015} or coexisting charge-density wave (CDW) orders \cite{Sooryakumar1980,Measson14,Grasset18}. In both cases, the Higgs mode energies are `pushed' in gap, and thus the mode becomes much more stable \cite{Cea2014Disorder,Podolsky11,Littlewood81,Littlewood82,Pekker15,Cea14}.

In this work, we show that long-range electron interactions present a way to manipulate these low-lying excitations and, in particular, to stabilise the Higgs mode. 
This proposal leverages the strong coupling between electrons and THz nanoplasmonic cavities \cite{Scalari12,Maissen14,Zhang16}
which allows for virtual scattering of external laser photons inside the cavity via two-photon diamagnetic processes and induces long-range, density-density electron interactions that are essentially unscreened and controlled by the laser parameters \cite{Gao2020-frank}. 
By choosing the laser to be red-detuned from the cavity resonance, we engineer 
long-range electron attraction on top of the intrinsic local attractions in a 2D BCS superconductor, which for simplicity we choose to be s-wave. 

By studying the fluctuations on top of the BCS ground state, 
we find that the long-range nature of the induced interactions qualitatively affects the collective excitations in the superconductor. 
We show that the induced long-range attraction allows for excitonic modes in an s-wave superconductor without competing intrinsic superconducting orders. 
More generally, this means that long-range attraction can enlarge the separation in energy between the quasiparticle continuum and all excitonic modes intrinsically present in the superconductor. 

Importantly, the Higgs mode is also pushed ``in gap" by long-range interactions; and the separation of its energy from the quasiparticle continuum depends linearly on the driving intensity. 
Consequently, the Higgs mode no longer decays through the quasiparticle excitations, substantially increasing its lifetime, which makes it easier to detect regardless of the experimental method \cite{Cea14}. We demonstrate the Higgs mode stabilisation with numerical simulations, comparing the oscillations of the superconducting gap with and without the long-range interactions following an initial excitation of the Higgs mode. 
Our work presents a novel mechanism for the stabilisation of the Higgs mode. 

 \begin{figure}[tbh]
     \centering
     \includegraphics[width=.85\linewidth]{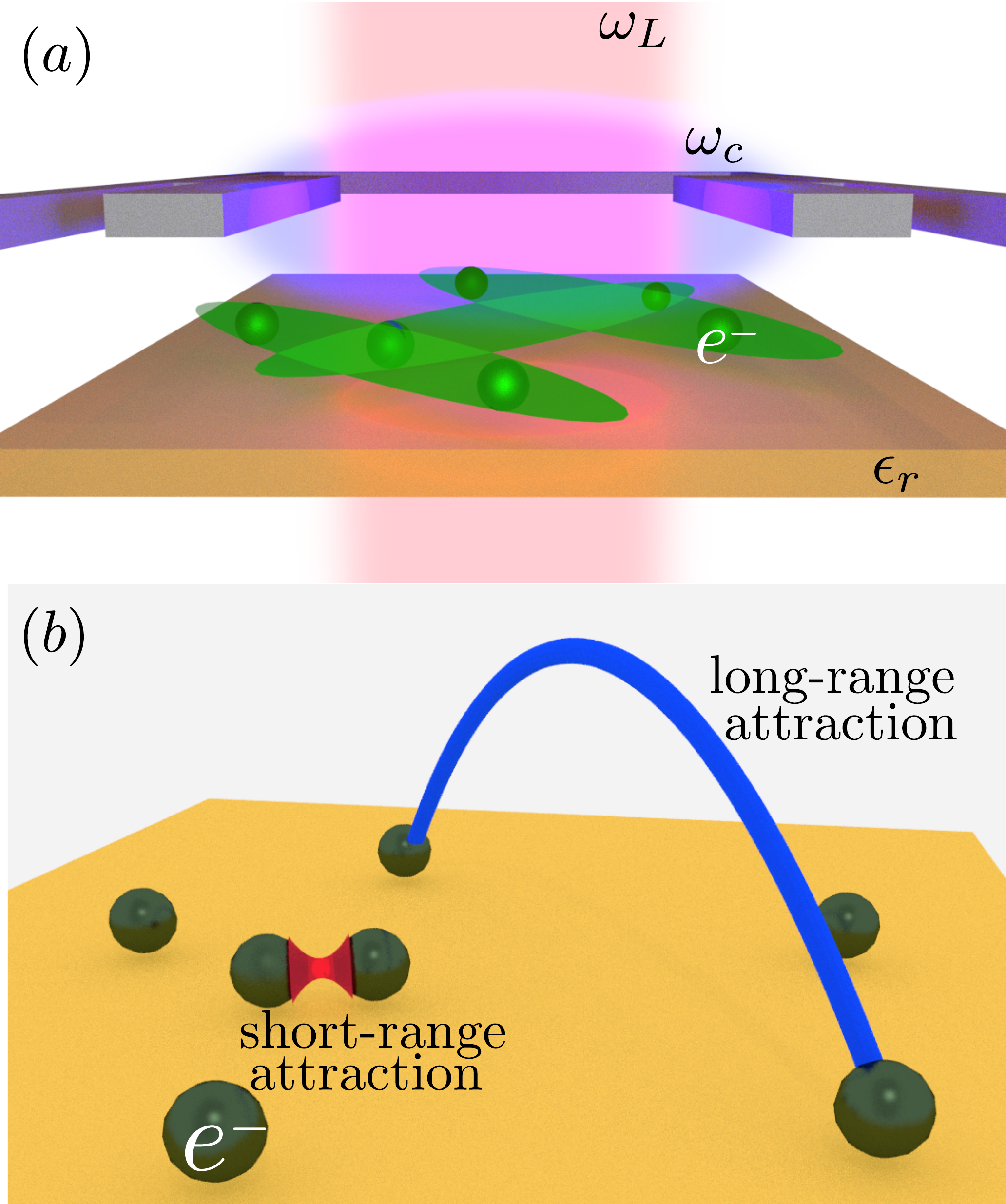}
     \caption{ (a) Setup: 
     electrons (spheres) in a 2D BCS superconductor (the green envelopes represent Cooper pairing) are evanescently coupled to a complementary split-ring cavity (the gray structure on top) \cite{Maissen14}. The blue shading symbolizes the cavity field. The coupled system is driven by a laser field (schematically shown by the red shading) with wavevector ${\bf q}_L $ in the vertical direction 
     and frequency $\omega_L$, which is red-detuned from the cavity frequency, $\omega_c$.
     (b) The coupling to the cavity and the driving fields result in long-range attractive interactions (blue arc) being induced on top of preexisting short-range attractions (red link).
     }
     \label{fig:setup}
 \end{figure} 
 
\section*{Results}
\paragraph*{\textbf{Setup.}}
Cavity-mediated long-range interactions have recently been considered in a number of different setups, covering both lattice and Bloch band electrons coupled to either an empty or a driven cavity \cite{Kiffner18,Schlawin19,Gao2020-frank,chakraborty2020,Li20,Li_eta_2020,Li_PRL_2020,chiocchetta2020,li2021effective,rokaj2021free}. 
Here, we consider the setup shown in Fig.~\ref{fig:setup}(a). 
A 2D superconductor lies in the $xy$-plane inside a substrate material with relative permittivity $\epsilon_r$. The relevant electrons of the superconductor are coupled to the evanescent field of a complementary split-ring cavity which sits on top of the substrate material, as e.g., in Refs.~\cite{Maissen14,Paravicini19}. The cavity is approximately described by a single-mode light field since the higher modes' frequencies are multiples of the fundamental mode frequency and are high above the frequency range of interest \cite{Gao2020-frank}. The coupling between the electrons and the cavity field is enhanced by the high degree of compression of the cavity mode volume 
$\mathcal{V}_c = \Lambda \lambda^3$ beyond the free-space limit \cite{Maissen14,Keller17}, where $\lambda$ is the wavelength of the cavity mode in the substrate material, and $\Lambda$ is the mode volume compression factor. 
Values of 
$\Lambda\approx 3\times 10^{-6}$ 
or even smaller have been reported in experiments and simulations \cite{Maissen14,Kim18}. 
Moreover, the cavity has been shown to be transparent at a broad range of frequencies except the narrow width centered on its resonance, $\omega_c$, where it is highly reflective \cite{Maissen14}. 
We assume the cavity mode lies along the $x$-direction and its field is polarised along the $y$-direction. 
The cavity is described by a vector potential
$
{\bf A}_\text{c} ({\bf r}) \propto \mathbf{e}_y \sqrt{1/\mathcal{V}_c  \epsilon_r \omega_c} \cos(q_0 x) \text{, }
$
where $\mathbf{e}_y$ is the unit vector in the $y$-direction, $q_0 = \omega_c \sqrt{\epsilon_r}/ c$, and $c$ is the speed of light in vacuum. 
The cavity-matter system is driven by a strong laser with intensity $I_\text{d}$ which produces an oscillating classical field with vector potential ${{\bf A}}_\text{d} (t) \propto \mathbf{e}_y\sqrt{I_{\text{d}}}\sin(|{{\bf q}}_L| z - \omega_L t)$, where ${\bf q}_L$ is the laser photon wavevector and $t$ is the time. The laser frequency, $\omega_L = \omega_c - \delta_c$, is red-detuned from the cavity resonance (i.e., $\delta >0$), such that the driving field interacts directly with the electronic system. 
We require the driving and cavity frequencies to be off-resonant from all transitions in the material and the substrate to minimise heating \cite{Gao2020-frank,Leggett11}.

\paragraph*{\textbf{Model.}}
In the Coulomb gauge, our BCS superconductor with local attractive interactions is described by the Hamiltonian (setting $\hbar = 1$)
\begin{equation}
H_{\text{mat}} = \sum\limits_{\ksig} \xi_{{\bf{k}}} c^\dagger_{\ksig} c_{\ksig} - \frac{V}{\mathcal{S}} \sum\limits_{{\bf{k}}_1,{\bf{k}}_2,{\bf{k}}_3} c^\dagger_{{\bf{k}}_1\uparrow} c^\dagger_{{\bf{k}}_2\downarrow} 
c_{{\bf{k}}_3\downarrow} c_{{\bf{k}}_1+{\bf{k}}_2-{\bf{k}}_3\uparrow} \text{,}
\label{eq: H SC mat}
\end{equation}
where $\xi_{\bf{k}}$ is the dispersion, measured from the chemical potential, $\mu$, of an electron in quasimomentum state ${\bf{k}}$, which can be populated with a spin-$\sigma$ electron by the action of the fermionic creation operator, $c^\dagger_{\ksig}$. $\mathcal{S}$ is the cavity area that the 2D superconductor couples to. For simplicity, we assume that the dispersion is rotationally symmetric (i.e., it depends only on $\abs{\mathbf{k}}$). 
The strength of the local (i.e., independent of $\mathbf{k}$) intrinsic attractive interactions is given by $V>0$. 
We also assume the usual cutoff energy for these interactions, $\epsilon_{\text{D}}$, which, in the case of phonon-mediated superconductivity, is identified with the Debye energy \cite{TinkhamSuperconductivity}. 

The coupling of the electrons to the cavity and the driving field induces effective long-range density-density interactions between the electrons that can be tuned attractive and remain essentially unscreened \cite{Gao2020-frank}.
Moreover, it has been shown that the direct heating of the electrons in this setup is insignificant \cite{Gao2020-frank}. In this work, we thus take an approach similar to that in Ref.~\cite{chiocchetta2020} and model the effect of the cavity and the driving by $H_{\text{long}} = - \sum_{{\bf{k}}_1,{\bf{k}}_2,{\bf{q}}} U_{\bf{q}}/\mathcal{S} \quad c^\dagger_{{\bf{k}}_1 + {\bf{q}},\uparrow} c^\dagger_{{\bf{k}}_2 -{\bf{q}},\downarrow} 
c_{{\bf{k}}_2,\downarrow} c_{{\bf{k}}_1,\uparrow}$, where
\begin{equation}
\frac{U_{\bf{q}}}{\mathcal{S}} = \frac{\alpha^2 }{\pi c^2 m^2 } \frac{\omega_c^2}{\omega_L^2} \frac{I_{\text{d}}}{  \Lambda \delta_c} \delta_{\pm {\bf q},q_0 \mathbf{e}_x} \text{.}
\label{eq: long range interaction strength}
\end{equation}
We remark that the attraction considered here is stronger for higher driving intensities and smaller detunings; however, $ \delta_c$ must be greater than relevant energy scales of the superconductor (the cavity-coupled plasmon energy \cite{Gao2020-frank} and the superconducting gap size \cite{Leggett11}) for us to be able to ignore retardation in the induced interactions. 
In the following, we study the zero-temperature properties of the model.

\paragraph*{\textbf{Zero-temperature gap structure.}}
Through the usual BCS reduction and mean-field (MF) 
decoupling of the full system ($H_{\text{mat}}+ H_{\text{long}}$), we obtain a gap equation 
which we evaluate at $T=0$K, 
\begin{equation}
    \Delta_{{\bf k}}
= \frac{V}{\mathcal{S}} \sum\limits_{{\bf k}^{'}} \frac{\Delta_{{\bf k}^{'}} }{ 2 E_{{\bf k}^{'}}} + \tilde{U} \frac{ \Delta_{{\bf k}} }{ E_{{\bf k}}} \text{,} \label{eq: gap eq}
\end{equation}
where 
$E_{{\bf k}} = \sqrt{ \xi_{{\bf k}}^2 + \Delta_{{\bf k}}^2}$ gives the Bogoliubov quasiparticle dispersion, which is measurable with ARPES \cite{DamascelliRMP2003}. $\tilde{U} = U_{q_0 \mathbf{e}_x}/\mathcal{S}$, and $\Delta_{{\bf k}} \equiv \sum_{{\bf k}'} (V/\mathcal{S} + 2\tilde{U}\delta_{{\bf k},{\bf k}'}) \langle c_{-{{\bf k}'}\downarrow} c_{{{\bf k}'}\uparrow} \rangle$ defines the MF. We note that we have also 
taken the $q_0 \rightarrow 0$ limit of 
$U_{{\bf q}}$ in the MF definition and in the gap equation \eqr{eq: gap eq} \cite{Schlawin19atom}.

We solve the gap equation and 
show in \fir{fig: T=0}(a) the zero-temperature gap functions 
in the radial direction around the Fermi surface with and without the induced long-range attractive interactions. 
In a usual BCS s-wave superconductor with only local electron attraction, the gap function $\Delta^{\text{in}} = 2 \epsilon_{\text{D}} \exp{-1/N(0)V}$ is a constant in $k$-space. Here $N(0)$ is the electron density of state per spin at the Fermi surface. 
In contrast, 
in \fir{fig: T=0}(a) we show that the presence of the long-range interactions causes the gap function to gain a structure in the radial direction. As for the azimuthal direction, the rotational symmetry of the electron dispersion ensures that the s-wave gap function has rotational symmetry in k-space. 
The gap is greatest at the (original) Fermi surface and flattens out further away ($|\xi_{{\bf k}}| \gtrsim \Delta^{T=0}_{k_F} $) from the Fermi surface. Importantly, the gap is larger everywhere in the presence of the long-range attractive interactions. In fact, for $\tilde{U} \ll 
\Delta^{\text{in}}$, 
the gap increases linearly with $\tilde{U}$: 
\begin{equation}
    \Delta_{{\bf k}}^{T=0} = \Delta^{\text{in}} + \left( \frac{\pi}{4} + \frac{\Delta^{\text{in}}}{\sqrt{(\Delta^{\text{in}})^2 + \xi^2_{{\bf k}}}} \right)\tilde{U}\text{,}
\end{equation}
where $\Delta^{\text{in}}$ is the gap size in the absence of 
the long-range interactions.
This is in contrast to the case of superconductors with long-range interactions only \cite{Yang00}. There, the gap function vanishes away from the Fermi surface ($|\xi_{{\bf k}}| > \tilde{U}$), where the gap size is $\tilde{U}$. The fact that in our system the gap size enhancement at $\abs{\mathbf{k}} = k_F$ is greater than $\tilde{U}$ highlights that the long-range interactions combine with the local interactions nonlinearly.

The enlargement of the gap function implies that the Bogoliubov quasiparticle excitation continuum is pushed to higher energies, reflecting an increase of superconducting critical temperature which we will discuss elsewhere. 
The mechanism at play here differs from the Eliashberg effect \cite{Wyatt66,Dayem67,Ivlev1973} and its quantum equivalent \cite{Curtis18}, which dynamically enhance the gap close to the critical temperature through quasiparticle redistribution \cite{Tikhonov18}, whereas in our case, the enhancement extends to zero temperature.

\begin{figure*}[t!]
    \centering
    \includegraphics[width=.85\linewidth]{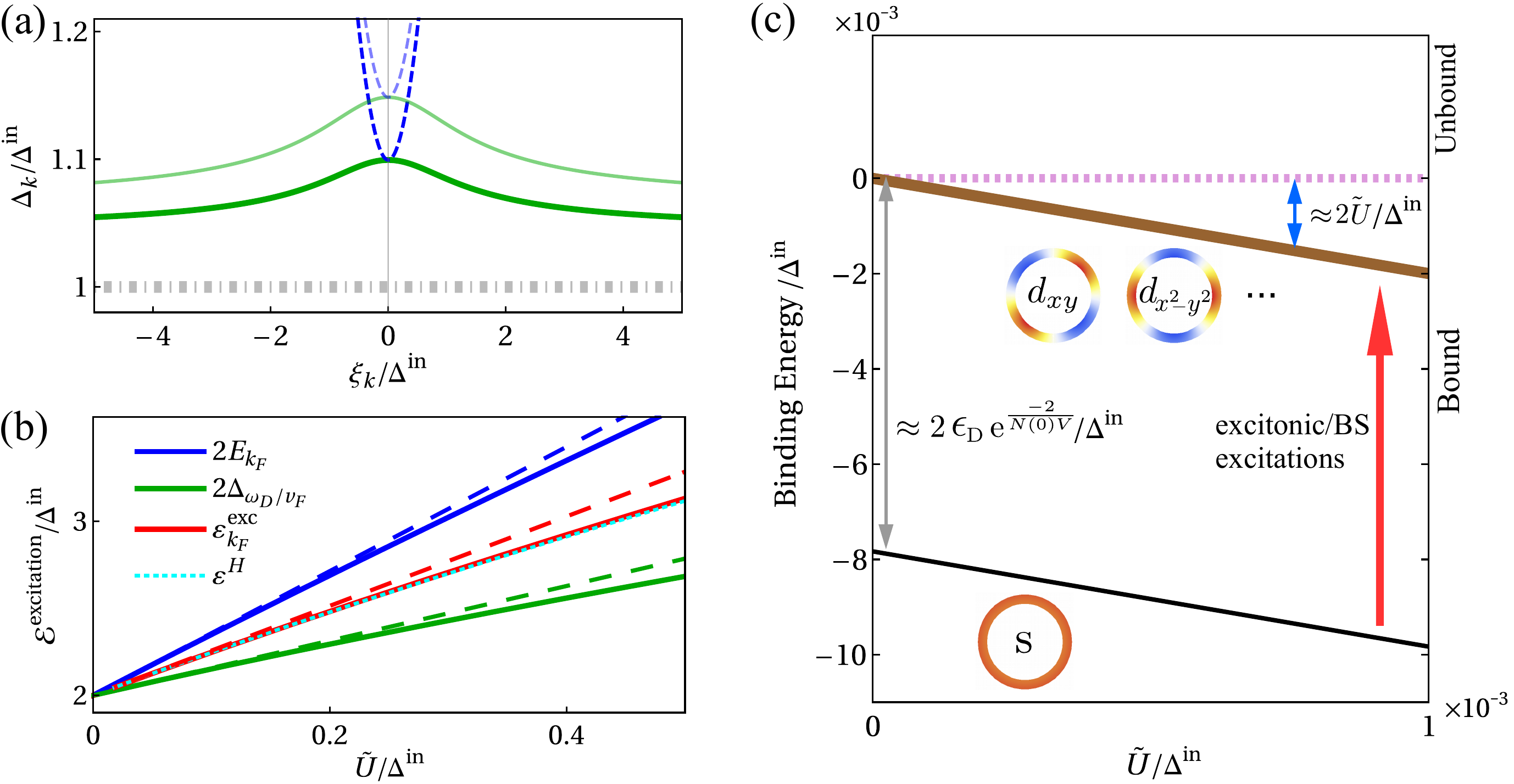}
    \caption{
    (a) The zero-temperature gap functions with two different long-range-interaction strengths are plotted along the radial direction in k-space with green lines. 
    The $x$-axis is rescaled by the Fermi velocity and normalised by the intrinsic gap size to show the electron dispersion $\xi_{\mathbf{k}} / \Delta^{\text{in}}$ at those wavevectors. 
    We also plot the Bogoliubov quasiparticle excitation energies for the system with dashed blue lines. 
    We use solid lines for $\tilde{U}/ \Delta^{\text{in}}\approx 0.056$, and semitransparent lines for $\tilde{U}/ \Delta^{\text{in}} \approx 0.085$. The gap function without the long-range interactions is plotted with a gray dash-dotted line. 
    (b) We plot the values of the various excitation energies of the system against $\tilde{U}$. $2E_{k_F}$ is the lowest Bogoliubov quasiparticle excitation energy, $\varepsilon^{\text{exc}}_{k_F}$ is the lowest excitonic excitation energy, $2\Delta_{\epsilon_{\text{D}}/v_F}$ is twice the smallest value the gap function takes, and $\varepsilon^{\text{H}}$ is the zero-momentum Higgs excitation energy.
    The dashed blue, red and green lines give the linear dependence of the corresponding energies on $\tilde{U}$ obtained from first-order perturbation theory. 
    The results are calculated for $T=0$K. In (a) and (b) we have used 
    $\epsilon_{\text{D}}/ \Delta^{\text{in}} = 113$. 
    (c) An energy diagram showing the electron-pair binding energy (from considering the Cooper problem) as a function of the long-range interaction strength. Without the long-range interactions ($\tilde{U}/\Delta^{\text{in}}=0$), the only one two-electron bound state is in the $s$-wave channel. Long-range interactions 
    cause states with higher angular momenta (e.g., $d$-wave) to be bound. 
    Cooper pairs in the condensate can be excited to these higher energy bound states to excite excitonic (a.k.a.\ Bardasis-Schrieffer) modes. 
    }
    \label{fig: T=0}
\end{figure*}

\paragraph*{\textbf{Collective modes.}}

To study the collective modes at zero centre-of-mass momenta, 
we go beyond BCS MF theory, taking Anderson's pseudospin approach set out in Refs.~\cite{Anderson58,Yang00}. 
With this approach, the MF Hamiltonian is mapped to a system of pseudospins interacting with pseudofields. The collective modes are represented as small pseudospin precessions around the ground state alignment (see Methods for more details). 

In this formalism, we find that excitonic excitations are supported. 
Their energy $\varepsilon^{\text{exc}}_{{\bf k}}$ (whose full expression is given in Methods) 
take their minimum value at the Fermi surface of the normal state. 
We show its value as a function of $\tilde{U}$ in \fir{fig: T=0}(b). 
We see that the excitonic excitations are in gap as expected. 
For $\tilde{U} \ll \Delta^{\text{in}}$, 
both the Bogoliubov and the excitonic excitation energies increase linearly in $\tilde{U}$. At $\abs{\mathbf{k}} = k_F$, 
\begin{align}
   2E_{k_F} &\approx 2\Delta^{\text{in}} + (2+\pi/2)\tilde{U} \text{, and}\\ 
   \varepsilon^{\text{exc}}_{k_F} &\approx 2\Delta^{\text{in}} + (1+\pi/2)\tilde{U} \text{.}
\end{align}
This means that the separation in energy between the excitonic excitations and the quasiparticle continuum can be tuned by the intensity of the external laser field. 

The cause for the appearance of these excitonic excitations can be understood from a discussion of the Cooper problem. In superconductors with only local attraction, there is only one two-electron bound state in the $s$-wave channel \cite{TinkhamSuperconductivity}. 
This is shown in \fir{fig: T=0}(c) at $\tilde{U}/\Delta^{\text{in}}=0$. 
However, as soon as the long-range interactions mediated by the cavity are present, more bound states, e.g., in $d$-wave channels, appear because the long-range interactions increase all binding energies by $\approx 2\tilde{U}$. 
This is shown on the right of \fir{fig: T=0}(c). We note that the original $s$-wave state's binding energy also increases due to the long-range interactions, and the $s$-wave state remains the most tightly bound. Thus, the Cooper pairs still condense in the $s$-wave state. 
A Cooper pair from the condensate, whose constituents have quasimomenta $\pm \mathbf{k}$, can now be excited to one of these more loosely bound states as opposed to becoming unbound. In the former case, we have an excitonic excitation with energy $\varepsilon^{\text{exc}}_{\mathbf{k}}$, while in the latter, a pair of quasiparticle excitations are produced, costing $2E_{\mathbf{k}}$ of energy. 
In superconductors with only local ($s$-wave) interactions, the electron pairs have no other bound state to scatter into, and thus no excitonic excitations are present. 
In the opposite limit, where superconductors only have very long-range attractive interactions, many excitonic modes are available, and the least energetic ones amongst them become gapless as the interaction range tends to infinity \cite{Yang00}.

Strikingly, the Higgs mode is also in gap in the presence of the long-range interactions even though the superconductivity is still $s$-wave and there is no disorder or coexisting charge-density-wave phase. The Higgs mode's energy, $\varepsilon^{\text{H}}$, equals to $\varepsilon^{\text{exc}}_{k_F}$ up to $\mathcal{O}(\tilde{U})$, 
but $\varepsilon^{\text{H}}$ is strictly below $\varepsilon^{\text{exc}}_{k_F}$ for finite $\tilde{U}$. 
This means that, just like with the excitonic modes, the external laser intensity controls the energy separation between the Higgs mode and the quasiparticle continuum. 
We discuss 
the inequality $\varepsilon^{\text{H}} < \varepsilon^{\text{exc}}_{k_F} $ in Methods. We also verified it numerically.  
In \fir{fig: T=0}(b), we show that for $\tilde{U} \ll \Delta^{\text{in}}$, 
\begin{equation}
    \varepsilon^{\text{H}} \approx \varepsilon^{\text{exc}}_{k_F} \text{.}
\end{equation}

This proximity in energy between the zero-momentum Higgs mode and the lowest excitonic mode in our relatively simple model is a general feature resulting from the long-range nature of the induced interactions. Therefore, it can serve as an indication that the superconductivity is supported by both short- and long-range interactions. 
We remark that in a more complicated model in which a subdominant superconducting channel is present, 
the excitonic modes corresponding to this channel can be lower in energy than the Higgs mode, as is the case in the absence of the long-range interactions. 
However, all other angular momentum channels which intrinsically do not sustain superconducting pairing support bound states in the presence of the long-range interactions. They give rise to a continuum of corresponding excitonic modes, and the edge of this continuum will be close to the Higgs mode energy. 

An important consequence of the Higgs mode 
being in gap is that it can no longer decay through quasiparticle excitations, which usually dampen the Higgs mode such that it decays as $t^{-1/2}$, where $t$ is time \cite{Volkov73,Cea14,Matsunaga14,TsujiPRB15}. 
To illustrate this point, we present numerically simulated order parameter oscillations following an excitation in the Higgs mode in \fir{fig:Higgs oscillation}. The numerics were performed by solving the equation of motion of the Anderson's pseudospins [see Method \eqr{eq: eq of motion S}]. 
In an experiment, the Higgs mode could be excited by a strong THz pulse centred at half the Higgs mode frequency, and its subsequent oscillations could be observed by measuring the transmittance of time-delayed THz probe pulses \cite{Matsunaga13, Matsunaga14,Vaswani2021,Buzzi19arxiv}. 
We see that without the long-range interactions, the oscillation decays quickly, in agreement with earlier studies \cite{Volkov73,Cea14,Matsunaga14,TsujiPRB15}. 
In contrast, in the presence of the long-range interactions, the gap shows clean oscillations. 
This way, the Higgs mode becomes a well-defined collective mode \cite{Cea14,Sun2020} when the long-range interactions are switched on, making it easier to detect and distinguish in this system \cite{Shimano20}. 
Since the Higgs mode and the excitonic modes have different symmetries, they could potentially be more clearly distinguished 
in Raman measurements \cite{Grasset18}. 

We also comment that the Nambu-Goldstone modes still manifest as plasmon oscillations. In the parameter regime considered in this work, the coupling with the cavity and driving modes softens the plasmons slightly (by an amount proportional to $\tilde{U}$) \cite{Gao2020-frank} as the long-range attraction has opposite signs to the Coulomb interactions.

\begin{figure}
    \centering
    \includegraphics[width=\linewidth]{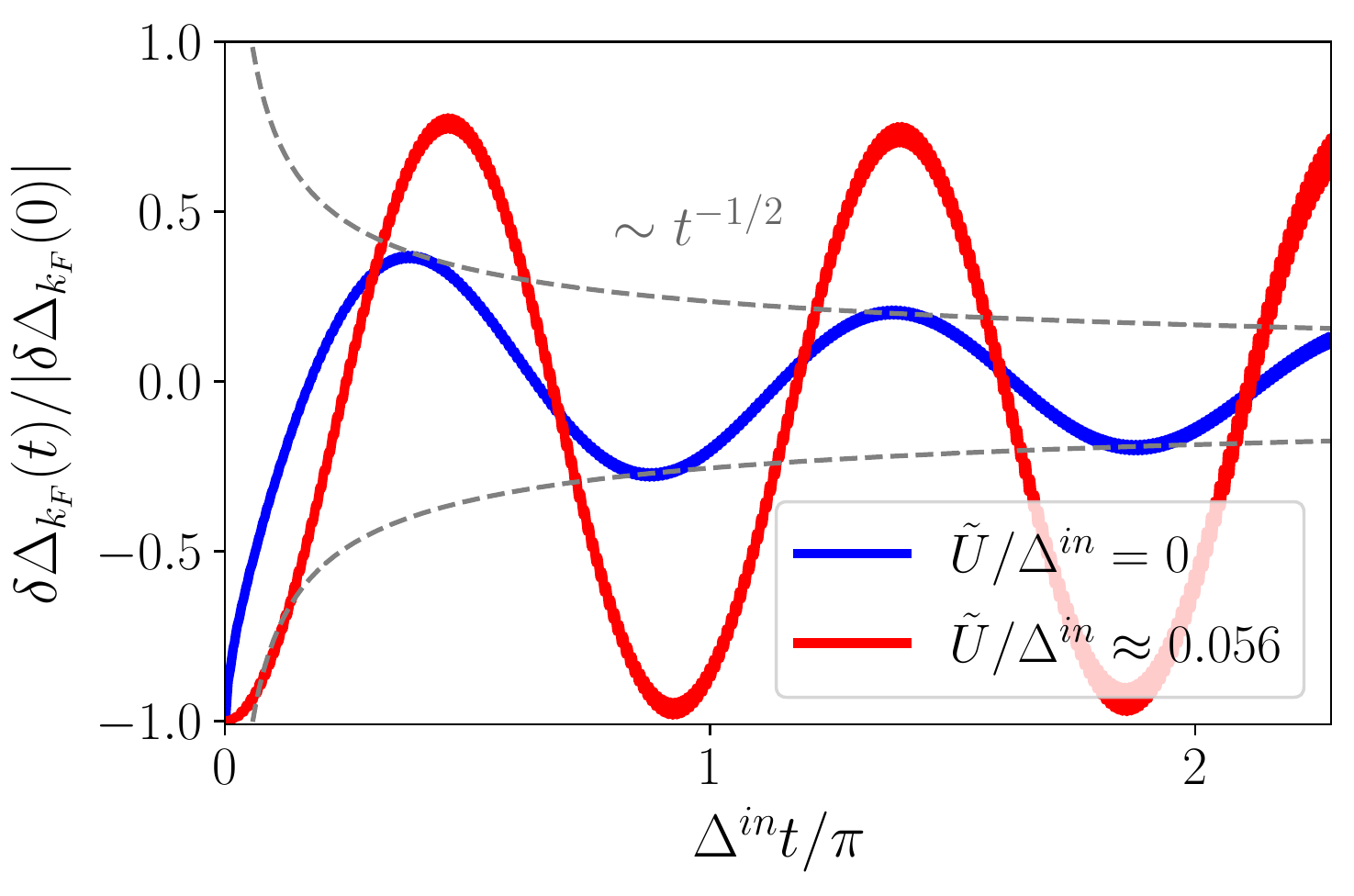}
    \caption{Normalised oscillations of the Higgs mode as a function of time following small perturbations of the superconducting gap with (in red) and without (in blue) the long-range interactions. Without the long-range interactions, the Higgs oscillation decays as $t^{-1/2}$ \cite{Cea14} as it is on the edge of the quasiparticle continuum. In contrast, in the presence of the long-range interactions, the Higgs oscillation does not decay. 
    The intrinsic parameters for the superconductor used here are the same as in \fir{fig: T=0}. 
    The initial perturbations of the order parameters are about $2$\% of their equilibrium values. 
    Due to the finite size of the initial perturbation, the gap oscillates around a value below the ground state value, hence the oscillations are offset by a small amount \cite{Yuzbashyan2005,Yuzbashyan2006,Schwarz2020}. The offset vanishes when the perturbation strength tends to zero, when the pseudospin equation of motion can be linearised. }
    \label{fig:Higgs oscillation}
\end{figure}

\section*{Discussion}

Our study shows that tunable cavity-mediated long-range interactions 
qualitatively alter the collective excitations at zero temperature in a superconductor which intrinsically only has short-range interactions. 
Two features are most notable. First, additional excitonic modes with finite energies appear in gap as the long-range interactions create additional electron-pair bound states. This also means that preexisting excitonic modes corresponding to intrinsic subdominant superconducting orders are further separated from the quasiparticle continuum, making them more distinguishable in experiments. 
Second, the Higgs mode of the system is also pushed below the quasiparticle excitation gap, and below the aforementioned excitonic modes. 
As a result, the Higgs mode becomes stable 
and easier to observe experimentally regardless of the detection process. 
We emphasise that what we present is one of the few known cases where the Higgs mode is below the quasiparticle continuum, and thus forms a well-defined collective mode. 
Moreover, the imposition of the long-range interactions does not necessarily alter the other characteristics (symmetries, coupling to light, etc.) of the Higgs mode. 
This makes long-range interactions useful 
for studying Higgs modes in superconductors when they are normally damped by quasiparticle excitations.

More generally, our study is also relevant to cold atomic gases, where short-range interactions are controlled using Feshbach resonances \cite{ChinRMP2010}, and long-range interactions can be engineered by coupling to optical cavities \cite{Landig2016,Camacho-Guardian17,Fan18,Schlawin19atom,Jaksch01,Munstermann00,Maschler05,Baumann2010,Mottl12,Zeytinoglu16,Mivehvar17,Kroeze18,Vaidya18,Norcia259,Guo19,Mivehvar19}. 
Collective modes of superfluids \cite{Randeria14,Combescot06,Diener08} and other phases of matter enabled by the long-range interactions in cold atomic gases \cite{Caballero_Benitez2015,Caballero_Benitez_2016,Landig2016} could be studied. 
For all these platforms, further insights into the low-energy collective modes are important towards understanding and controlling non-equilibrium dynamics of these coupled light-matter systems. 

For future work, as mentioned above, the role of long-range attraction in enhancing the critical temperature of conventional superconductors is still to be elucidated. 
In this regard, recent theoretical studies on similar cavity-coupled electron systems have shown that electron-phonon interactions can be enhanced by coupling to the cavity \cite{Sentef18,Hagenmuller2019}, though this does not translate into an enhancement of the critical temperature. 
It would be desirable to derive a (possibly extended) Ginzburg-Landau (GL) theory to describe superconductors with both short- and long-range interactions phenomenologically close to the critical temperature. 
Such a theory could describe spatial inhomogeneity and yield results on measures, such as correlation lengths, that characterise a superconductor . 
Moreover, the question of how cavity-mediated interactions can modify non-BCS superconductivity remains to be addressed. Studies in this direction could help to shed light on the reported strong enhancement of superconductivity by strong coupling to a cavity \cite{Thomas19b}.

\section*{Methods}

\paragraph*{\textbf{Cooper's problem.}}
The wavefunction of the Cooper pair has the form \cite{TinkhamSuperconductivity}
\begin{equation}
    \ket{\Phi} = \sum\limits_{0<\xi_{\mathbf{k}} <\epsilon_{\text{D}}} g(\mathbf{k}) c^\dagger_{\mathbf{k}\uparrow} c^\dagger_{-\mathbf{k}\downarrow} \ket{\text{FS}} \text{,}
\end{equation}
where $\ket{\text{FS}}$ represents the Fermi sea at $T=0$K. By substituting this into the full system Hamiltonian ($H_{\text{mat}}+H_{\text{long}}$) and taking the $q_0 \rightarrow 0$ limit, we find that $g(\mathbf{k})$ satisfies 
\begin{equation}
    E g(\mathbf{k}) = 2\xi_{\mathbf{k}} g(\mathbf{k}) - \sum_{\mathbf{k}'} (\tilde{V} + 2\tilde{U}\delta_{{\bf k},{\bf k}'}) g(\mathbf{k}') \text{,}
\end{equation}
where $E$ is the two-electron state energy and $\tilde{V} = V/\mathcal{S}$. We rearrange the terms to get
\begin{equation}
    \left( E- 2\xi_{\mathbf{k}} + 2 \tilde{U}\right) g(\mathbf{k}) = \tilde{V} \sum_{\mathbf{k}'} g(\mathbf{k}') \text{,}
\end{equation}
where the RHS is independent from $\mathbf{k}$. 
For the s-wave two-electron state, the RHS is nonzero. 
We follow the usual procedure of cancelling out common factors from both sides and perform the integration in $\mathbf{k}$-space \cite{TinkhamSuperconductivity} to obtain the s-wave binding energy
\begin{equation}
    E - 2\xi_{\mathbf{k}} \approx - 2\tilde{U} - 2\epsilon_{\text{D}} \ee^{\frac{-2}{N(0)V}} \text{.}
\end{equation}
For the two-electron states with higher angular momenta, the RHS is zero as the intrinsic attraction has no component in any higher angular momentum channels. Therefore, the binding energy in a higher momentum channel is $-2\tilde{U}$.
These results are as shown in \fir{fig: T=0}(c). 

\paragraph*{\textbf{Anderson's pseudospin formalism.}}
We study the BCS reduced Hamiltonian and investigate the collective modes in BCS superconductors at $T=0$K. This problem is more conveniently represented by mapping the electron operators to pseudospin operators \cite{Anderson58,Yang00},
\begin{align}
    S_{{\bf k}}^z =& (c^\dagger_{{\bf k},\uparrow} c_{{\bf k},\uparrow} + c^\dagger_{-{\bf k},\downarrow} c_{-{\bf k},\downarrow} - 1)/2\\
    S_{{\bf k}}^+ =& c^\dagger_{{\bf k},\uparrow} c^\dagger_{-{\bf k},\downarrow} \\
    S_{{\bf k}}^- =& c_{-{\bf k},\downarrow} c_{{\bf k},\uparrow} \text{.}
\end{align}
The BCS MF Hamiltonian in this representation is
\begin{align}
    H_{\text{MF}} =& - \sum\limits_{{\bf k}} \mathbf{B}_{{\bf k}} \cdot \mathbf{S}_{{\bf k}} \text{, where} \label{eq: H pseudospin}\\
    \mathbf{B}_{{\bf k}} =& -2 \xi_{{\bf k}} \mathbf{e}_z + 
    2 \sum\limits_{\mathbf{k}'} (\tilde{V} + 2 \tilde{U}\delta_{\mathbf{k},\mathbf{k}'}) \langle \mathbf{S}^{\perp}_{\mathbf{k}'} \rangle
    \text{,}
\end{align}
and $\mathbf{S}^{\perp}_{\mathbf{k}}$ refers to the component of $\mathbf{S}_{\mathbf{k}}$ in the $xy$-plane. 
For the ground state,
\begin{equation}
    \mathbf{B}^{0}_{{\bf k}} = -2 \xi_{{\bf k}} \mathbf{e}_z + 2 \Delta_{{\bf k}} \mathbf{e}_x \text{.}
\end{equation}
Here, we have assumed that $\mathbf{B}^{0}_{{\bf k}}$ lies in the $xz$-plane without loss of generality. 
We then go beyond the ground state and consider perturbations $\delta \mathbf{S}_{{\bf k}} = \mathbf{S}_{{\bf k}} - \mathbf{S}^0_{{\bf k}} $ on top of the MF ground state. Here, $\mathbf{S}^0_{{\bf k}}$ is the expectation value of $\mathbf{S}_{{\bf k}}$ in the MF ground state. 
From \eqr{eq: H pseudospin}, we obtain the equation of motion for $\mathbf{S}_{\mathbf{k}}$,
\begin{equation}
    \dv{\mathbf{S}_{{\bf k}}}{t} = \mathbf{S}_{{\bf k}} \times \mathbf{B}_{{\bf k}} \text{.} \label{eq: eq of motion S}
\end{equation}
We linearise it to get the equation of motion for $\delta \mathbf{S}_{{\bf k}}$,
\begin{equation}
   \dv{\delta \mathbf{S}_{{\bf k}}}{t} = \delta \mathbf{S}_{{\bf k}} \times \mathbf{B}^{0}_{{\bf k}} + \mathbf{S}_{{\bf k}} \times \delta \mathbf{B}_{{\bf k}} \text{,}
\end{equation}
where $\delta \mathbf{B}_{{\bf k}} = \mathbf{B}_{{\bf k}} -\mathbf{B}^{0}_{{\bf k}}$. 
We re-formulate this as an eigenvalue problem: 
\begin{align}
    & \omega^2 \phi_{{\bf k}} \nonumber\\
    =& \sum\limits_{{\bf k}'} \mathcal{M}_{{\bf k k}'} \phi_{{\bf k}'} \nonumber\\
    =& \left( B^0_{{\bf k}} \right)^2 \phi_{{\bf k}} - B^0_{{\bf k}} \sum\limits_{{\bf k}'} (\tilde{V} + 2\tilde{U} \delta_{{\bf k},{\bf k}'}) \phi_{{\bf k}'} \nonumber\\
    -& \cos{\theta_{{\bf k}}} \sum\limits_{{\bf k}'} (\tilde{V} + 2\tilde{U} \delta_{{\bf k},{\bf k}'}) B^0_{{\bf k}'} \cos{\theta_{{\bf k}'}} \phi_{{\bf k}'} \nonumber\\
    +& \cos{\theta_{{\bf k}}} \sum\limits_{{\bf k}',{\bf k}''} (\tilde{V} + 2\tilde{U} \delta_{{\bf k},{\bf k}'}) (\tilde{V} + 2\tilde{U} \delta_{{\bf k}',{\bf k}''}) 
    \cos{\theta_{{\bf k}'}} \phi_{{\bf k}''} \text{,}
    \label{eq: excitation eigenproblem}
\end{align}
where $S^y_{{\bf k}} \propto \phi_{{\bf k}} $, $\tan{\theta_{{\bf k}}} = -\Delta_{{\bf k}}/\xi_{{\bf k}} $ and $B^0_{{\bf k}} = | \mathbf{B}^0_{{\bf k}} |$. 

For excitonic modes the perturbations $\Phi^{\text{exc}}_{k}=( ..., \phi^{\text{exc}}_{{\bf k}},... )$ are local in each energy shell. $\Phi^{\text{exc}}_{k}$ satisfies $ \phi^{\text{exc}}_{{\bf k}} = 0 \quad \forall |{\bf k}| \neq k $ and $ \sum_{|{\bf k}|=k} \phi^{\text{exc}}_{{\bf k}} =0$. The energy of the excitonic modes index by $\mathbf{k}$ is
\begin{equation}
    \varepsilon^{\text{exc}}_{{\bf k}} = \sqrt{ \left( B^0_{{\bf k}} \right)^2 - 2 B^0_{{\bf k}} \tilde{U} \left( 1 + \cos^2\theta_{{\bf k}} \right) + 4 \tilde{U}^2 \cos^2\theta_{{\bf k}} } \text{.}
\end{equation}
The energy is lowest on the Fermi surface, $\varepsilon^{\text{exc}}_{k_F} = 2\sqrt{\Delta_{k_F}^2 - \tilde{U} \Delta_{k_F}}$. We emphasise that this energy is still higher than twice the superconducting gap without the long-range interactions.

The rest of the eigen-modes are rotationally symmetric in $k$-space. The Nambu-Goldstone mode appears as a zero-energy mode that is symmetric about the Fermi surface and satisfies $\phi^{\text{NG}}_{{\bf k}} \propto \Delta_{{\bf k}}$. This condition confirms that the Nambu-Goldstone mode indeed corresponds to the phase fluctuation of the order parameter. Through the Anderson-Higgs mechanism, this mode is absorbed into the longitudinal component of the gauge field and appears as plasmon oscillations \cite{Anderson58}. 

The next lowest energy-mode is the zero-momentum Higgs mode of amplitude fluctuations. The mode function, $\Phi^{\text{H}}$, is antisymmetric about the Fermi surface. We obtain its excitation energy numerically, using the antisymmetry of the mode function. 
We show analytically that this Higgs excitation energy is lower than the excitonic excitation energy by plugging an approximate mode function (inspired by the Higgs mode function in superconductors with only local electron attractions), $\phi^{\text{ap}}_{{\bf k}} \propto \Delta_{{\bf k}} / \lbrack(E_{{\bf k}} - \tilde{U}) \xi_{{\bf k}} \rbrack $ into \eqr{eq: excitation eigenproblem}, 
\begin{equation}
    \sum\limits_{{\bf k}'} \mathcal{M}_{{\bf k k}'} \phi^{\text{ap}}_{{\bf k}'}  = \left(4\Delta^2_{{\bf k}} - 4\tilde{U}\frac{\Delta^2_{{\bf k}}}{E_{{\bf k}}}\right) \phi^{\text{ap}}_{{\bf k}} \leq (\varepsilon^{\text{exc}}_{k_F})^2 \phi^{\text{ap}}_{{\bf k}} \text{.}
\end{equation}
This shows that the true zero-momentum Higgs mode function results in an energy lower than $\varepsilon^{\text{exc}}_{k_F}$ for finite attractive long-range interactions. 

In \fir{fig: T=0}(b), we showed that $\varepsilon^{\text{H}}$ is very close to $\varepsilon^{\text{exc}}_{k_F}$. Here, we note that this difference grows with the strength of the long-range interactions. 
In addition, we note that the range of the induced interactions is, though long, still finite in reality. The finite range slightly weakens the pairing in higher angular momentum channels relative to the s-wave channel \cite{Schlawin19atom}, thus we should expect the separation between the Higgs mode and the excitonic modes to be marginally larger.

\paragraph*{Acknowledgements}
This work has been supported by the European Research Council under the European Union's Seventh Framework Programme (FP7/2007-2013)/ERC Grant Agreement No.\ 319286 Q-MAC and by EPSRC grant No.\ EP/P009565/1. F. S. acknowledges support from the Cluster of Excellence `Advanced Imaging of Matter' of the Deutsche Forschungsgemeinschaft (DFG) - EXC 2056 - project ID 390715994.

\putbib
\end{bibunit}
\end{document}